\documentclass[floatfix,aps,prd,superscriptaddress,twocolumn,preprintnumbers]{revtex4-1}

\usepackage{float}
\usepackage{amsmath,amssymb,bm,bbm}
\usepackage{graphicx,graphics,color}
\usepackage[colorlinks=true,linkcolor=blue,citecolor=blue,urlcolor=blue]{hyperref}
\usepackage{dsfont}
\usepackage{slashed}
\usepackage{mathtools}
\usepackage{feynmp-auto}
\usepackage{tikz}
\usepackage{tabularx}
\usepackage[utf8]{inputenc}
\usepackage{ulem}
\usepackage{multirow}
\usepackage{dcolumn}
\usepackage{soul}
\usepackage{graphicx}%
\usepackage{braket}
\usepackage{chngcntr}

\newcommand{\pv}{{\bf p}}

\newcommand{\qv}{{\bf q}}

\newcommand{\kv}{{\bf k}}
\newcommand{\bv}{{\bf b}}

\begin{document}
\title{Probing nonperturbative transverse momentum dependent PDFs with chiral perturbation theory: the $\bar{d}-\bar{u}$ asymmetry}

\author{Marston Copeland}

\author{Thomas Mehen}

\affiliation{\mbox{Department of Physics, Duke University, Durham, North Carolina 27708, USA}}

\begin{abstract}
%We perform the first phenomenological study of TMD PDFs in $\chi$PT. In this analysis, we review the formalism used for studying the collinear $\bar{d}-\bar{u}$ asymmetry in the proton PDFs and identify a key power counting issue with the formalism. After developing a novel prescription for alleviating the issue, we match the TMD PDFs onto chiral EFT in order to predict a “TMD $\bar{d} - \bar{u}$ asymmetry” in the proton, which can be compared with future experiments.

We use chiral perturbation theory to study the long distance regime of transverse momentum dependent parton distribution functions (TMD PDFs). Chiral corrections to the TMD PDFs are computed from proton to pion/baryon splittings. For consistent power counting, we find that the fraction of the proton's momentum that a pion may carry must be kept small. We make predictions for a $\bar{d}-\bar{u}$ asymmetry in the proton's TMD PDFs and find that the effective theory gives a natural exponential suppression of the TMD PDF at long distances. We then explore the effects that additional nonperturbative physics may have on the TMD $\bar{d}-\bar{u}$ asymmetry.

%We discuss the importance of the region of integration for the pion's momentum fraction in order to preserve the chiral expansion. We illustrate these effects by comparison dimensionally regulated calculations against new SeaQuest data and find excellent agreement. We then apply these techniques to recent calculations of TMD hadronic distribution functions in chiral effective theory. We use these results to make predictions for the TMD $\bar{d}-\bar{u}$ asymmetry in the proton. 
\end{abstract}

\date{\today} 
\maketitle

%%%%%%%%%%%%%%%%%%%%%%%%%%%%%%%%%%%%%%%%%%%%%%%%%%%%%%%%%%%%%%%%%%%%%%%%%%%

The $\bar{d}-\bar{u}$ asymmetry \cite{Thomas:1983fh, HERMES:1998uvc, NuSea:2001idv} is an important probe of nonperturbative effects, like chiral symmetry breaking, in the proton's structure. Many nonperturbative models have been used to describe the asymmetry with reasonable success \cite{Schreiber:1991tc,Kumano:1997cy,Chang:2011vx,Alberg:2012wr}, but a description rooted in QCD is most straightforwardly made using chiral perturbation theory ($\chi$PT) \cite{Scherer:2002tk,Scherer:2009bt}.

In $\chi$PT, an effective field theory (EFT) of QCD, the proton's structure gains long-distance contributions from fluctuations into intermediate meson and baryon states. In this picture, the $\bar{d}-\bar{u}$ asymmetry can be understood by the proton's likeliness to fluctuate to a $\pi^+n$ state over a state containing a $\pi^-$. The relationship between the proton's parton distribution function (PDF) and chiral operators describing the fluctuations (sometimes called hadronic splitting functions \cite{Burkardt:2012hk, Salamu:2014pka, Salamu:2018cny,Salamu:2019dok} or hadronic distribution functions \cite{Copeland:2024wwm}) was established using EFT in Ref. \cite{Chen:2001nb}. In this framework, a quark PDF in the proton is described by a convolution between the valence quark PDFs in an intermediate hadron state and the chiral hadronic distribution functions. 

These ideas have been applied to study collinear parton distribution functions \cite{Chen:2001nb, Arndt:2001ye, Burkardt:2012hk, Shanahan:2013xw, Ji:2013bca, Salamu:2014pka, Salamu:2018cny, Salamu:2019dok, Wang:2022bxo}, generalized parton distributions \cite{Ando:2006sk, Moiseeva:2012zi, He:2022leb, Strikman:2009bd, Gao:2024ajz}, and light-cone distribution functions \cite{Chen:2003fp}. A number of $\chi$PT calculations of the $\bar{d}-\bar{u}$ asymmetry have been compared with the NuSea E866 experiment \cite{NuSea:2001idv}, showing excellent agreement with the measurements \cite{Salamu:2014pka,Salamu:2019dok,Barry:2018ort, Alberg:2021nmu}. The majority of these calculations use relativistic chiral EFT and a variety of regularization prescriptions for the divergent pion loops that appear in the calculation. These prescriptions include hard cutoffs \cite{Burkardt:2012hk, Salamu:2014pka}, Pauli-Villars \cite{Barry:2018ort},  and dipole regulators \cite{Alberg:2021nmu, Salamu:2019dok}. While these calculations are phenomenologically very successful, they are dependent on the regulator parameters which are left finite in the calculations in order to describe the data. Regulating the $\chi$PT calculations with dimensional regularization instead would produce a cutoff independent result that also naturally preserves the symmetries of the theory. However, to our knowledge, a comparison with the $\bar{d}-\bar{u}$ measurements using dimensional regularization is absent from the literature. 
%However, despite the popularity of dimensional regularization in particle physics calculations, to our knowledge, a comparison with the $\bar{d}-\bar{u}$ measurements using dimensional regularization is absent from the literature. An analysis of the $\bar{d}-\bar{u}$ asymmetry in chiral effective theory using dimensional regularization is one goal of this paper.

Recently, some progress has been made for matching transverse momentum dependent PDFs (TMD PDFs) onto $\chi$PT as well \cite{He:2019fzn, Copeland:2024wwm}. In Ref. \cite{Copeland:2024wwm} the authors argue that, like in the collinear formalism, the TMD PDF can be expressed as convolution between a TMD hadronic distribution function and the valence quark TMD PDF in the intermediate hadron. This is useful because, historically, the large $\bv_T$ regime for the TMD PDF has been theoretically unreachable and usually must be described with some nonperturbative model \cite{Boussarie:2023izj,Signori:2013mda, Anselmino:2013lza,Scimemi:2019cmh,Bertone:2019nxa}. However, the hadronic corrections in $\chi$PT provide an avenue to systematically study corrections to the TMD PDF at long distances, $\bv_T \sim {\cal O}(1/m_\pi)$. 
%Historically, the large $\bv_T$ regime for the TMD PDF has been theoretically unreachable and usually must be described in an ad. hoc. manner with some nonperturbative model. 

In this letter, we initiate the first phenomenological study of unpolarized TMD PDFs in $\chi$PT. In analogy with collinear PDFs, we aim to predict the $\bar{d}-\bar{u}$ asymmetry for TMD PDFs in the proton. Our TMD hadronic distribution function results are cutoff independent meaning we have limited flexibility tuning parameters to describe data. Therefore, we revisit the $\bar{d}-\bar{u}$ asymmetry for collinear PDFs in $\chi$PT, using dimensional regularization instead of other regulators, to understand how to describe experiment without a cutoff parameter. In this analysis, we find that it is essential to restrict the fraction of longitudinal momentum carried by the pion in the proton to small values. We compare our collinear calculations for the asymmetry to the NuSea E866 \cite{NuSea:2001idv} and the SeaQuest E906 data \cite{FNALE906:2022xdu} and find excellent agreement. We then apply this prescription to the TMD hadronic distribution functions \cite{Copeland:2024wwm} and make predictions for a $\bar{d}-\bar{u}$ asymmetry in the proton's TMD PDFs. We compare our results against perturbative QCD (pQCD) calculations and find that $\chi$PT provides a suppression at large $\bv_T$ that pQCD cannot replicate. Finally, we explore the possibility that there are additional nonperturbative effects not described by the chiral theory and we illustrate how such effects would be distinguished from our calculations. 

We begin by revisiting the $\bar{d}-\bar{u}$ asymmetry for collinear PDFs in the proton. To study this observable, we will make use of the convolution formalism introduced in Ref. \cite{Chen:2001nb}. Generally, the collinear PDFs can be matched onto chiral operators using the following expression,
\begin{equation}
    f_{q/p}(x,\mu) = \sum_H \int_x^1 \frac{dy}{y}q^v_H\bigg(\frac{x}{y};\frac{\mu}{\Lambda_\chi}\bigg)f_{Hp}\bigg(y;\frac{m_\pi}{\Lambda_\chi}\bigg)
\label{eq: col conv}
\end{equation}
where $q^{v}_H(x) = q_H(x) - \bar{q}_H(x)$ is the valence PDF in the intermediate hadron \cite{Chen:2001nb, Salamu:2014pka, Salamu:2019dok}. $f_{Hp}(x)$ is the hadronic distribution function (HDF) in the proton. From an EFT point of view, the valence PDFs, $q^v_H$, are high-energy Wilson coefficients, meaning they contain physics at the scale $\mu$, which is at or above the large scale, $\Lambda_\chi \sim 1$ GeV \cite{Chen:2003fp, Lin:2020ssv}. In fact, for the convolution in Eq. (\ref{eq: col conv}), the $q^v_H$ are defined in the limit where $p_\pi \sim m_\pi = 0$ since $m_\pi \ll \Lambda_\chi$.
On the other hand, the HDFs, $f_{Hp}$, are defined in terms of effective theory operators and hence they contain all of the low-energy physics at the scale $m_\pi$. 

\begin{figure}[h!]
    \centering
    \includegraphics[width=\linewidth]{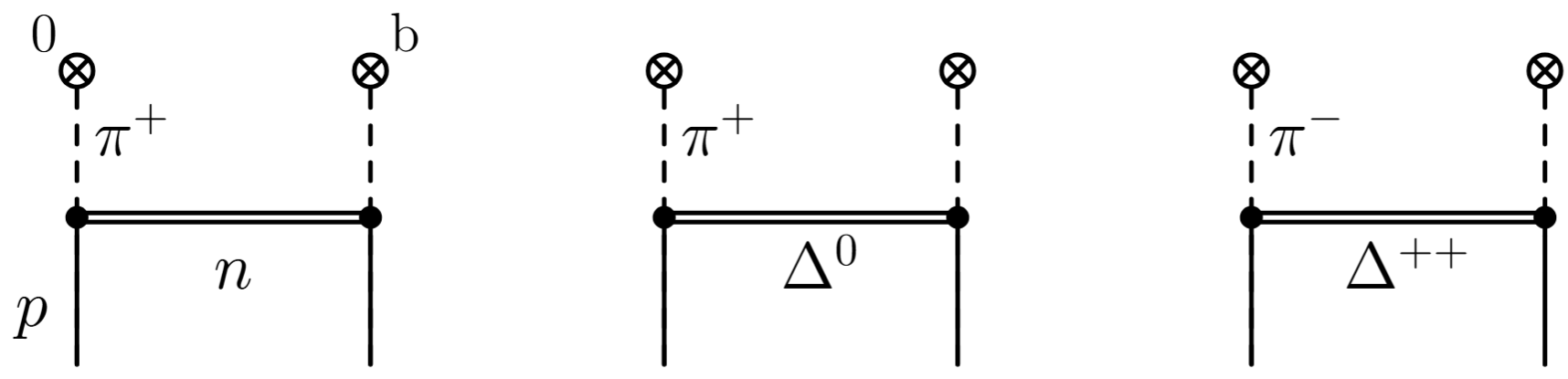}
    \caption{Feynman diagrams for the one loop contributions to $f_{\pi p}(y)$. Contributions from the $p\to \pi^+n$ splitting are depicted by the first diagram, while contributions from $p\to \pi^+\Delta^0$ and $p\to \pi^-\Delta^{++}$ are given by the second diagrams.}
    \label{fig: Delta loops}
\end{figure}

For the $\bar{d}-$$\bar{u}$ asymmetry, only the distribution of $\pi^+ - \pi^-$ states in the proton is relevant. This distribution is defined by,
\begin{equation}
    f_{\pi p}(y) = \int \frac{db^-}{2\pi} e^{-iy P^+b^-} \bra{p} {\cal O}_\pi(b^-,0) \ket{p}
\label{eq: fpi col}
\end{equation}
where the operator dependent on some position, $b$, is,
\begin{equation}
\begin{aligned}
    {\cal O}_\pi(b, 0) = &\frac{f_\pi^2}{8}{\rm Tr}\big[\Sigma^\dagger(b) \tau^a (i n\cdot \partial) \Sigma(0) \\
    &+ \Sigma(b) \tau^a (i n\cdot \partial) \Sigma^{\dagger}(0)].
\end{aligned}
\label{eq: Opi}
\end{equation}
At leading order, the operator receives corrections from $p\to \pi^+ n$ interactions. It is also well known that the $\Delta$ resonance plays a substantial role for nucleon structure in $\chi$PT \cite{Pascalutsa:2005nd,Scherer:2009bt,Scherer:2012xha, Salamu:2014pka}. %This is because the $\Delta-$nucleon mass splitting, $\delta = M_\Delta - M_N$ = 293 MeV, is a mass scale not much larger than the pion mass and hence needs to be incorporated in the effective theory. 
With the $\Delta$ resonance included, the $p \to \pi^+ \Delta^0$ and $p \to \pi^- \Delta^{++}$ interactions also contribute to $f_{\pi p}(y)$ at leading order. The diagrams we need to calculate are shown in Fig. \ref{fig: Delta loops}. We calculate the contributions using dimensional regularization and a minimal subtraction scheme. The results are given in Appendix \ref{app: HDF results}.

\begin{figure*}[htp]
    \centering
    \vspace{-.5cm}
    \includegraphics[width=.65\linewidth]{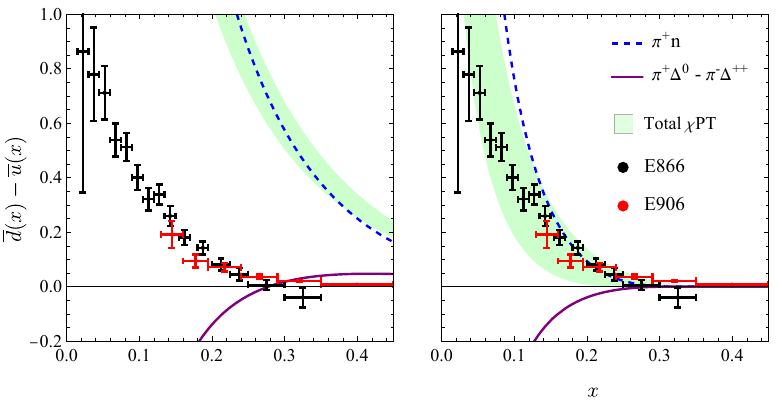}
    \caption{Comparison of $\chi$PT calculations with NuSea E866 \cite{NuSea:2001idv} and SeaQuest E906 data \cite{FNALE906:2022xdu} before (left) and after (right) enforcing chiral power counting in the pion's momentum fraction. The uncertainty on the left figure stems from varying the result by ${\cal O}(m_\pi/\Lambda_\chi)$ in order to estimate the size of subleading effective theory corrections. The uncertainty on the second diagram comes from varying the $y_{max}$ from $0.25 \le y_{max} \le 0.4$. The central values of the $p\to\pi^+n$ and $p\to \pi\Delta$ contributions are shown by the dotted blue and solid purple lines, respectively. The pion PDFs \cite{Clark:2016jgm} are evaluated at the scale $\mu^2 = 54$ GeV$^2$ to compare with the E866 data. }
    \vspace{-.25cm}
    \label{fig: col. d-u}
\end{figure*}

Naively, the convolution given in Eq. (\ref{eq: col conv}) encourages an integration over the full range of the pion's momentum fraction in the proton, $y$, from 0 to 1.  However, as demonstrated in Fig. \ref{fig: col. d-u}, a straightforward application of this formalism fails catastrophically. In fact, the theoretical prediction using dimensional regularization is $10-15$ times greater than the observed experimental data. 

This formalism fails because the convolution in Eq. (\ref{eq: col conv}) unintentionally allows the soft pions to carry large momentum scales, which should be forbidden by $\chi$PT. To illustrate what is happening, consider the delta function that constrains $y$,
\begin{equation}
    \delta(k^+ - yP^+),
\label{eq: delta 1}
\end{equation}
which naturally appears in the HDF caluclations \cite{Burkardt:2012hk,Chen:2001nb, Salamu:2014pka,Copeland:2024wwm}. Here, $k^+$ is the pion's lightcone momentum and $P^+$ is the nucleon's large lightcone momentum. For momentum fraction values $y \gg m_\pi/M$, where $M$ is the mass of the proton, $y P^+$ is formally a large quantity with respect to the soft pion momentum, $k^+$. In this region the delta function in Eq. (\ref{eq: delta 1}) violates $\chi$PT power counting and must be properly expanded. Treating $k^+$ as small with respect to $yP^+$, we should instead have the delta function
\begin{equation}
    \delta(k^+ - y P^+) \to \delta(y P^+) + {\cal O}(k^+).
\label{eq: expanded delta}
\end{equation}
Interestingly, this delta function vanishes because $yP^+ \gg m_{\pi} > 0$.

We argue that for $\chi$PT power counting to be preserved, one must match onto the original operators in Eq. (\ref{eq: fpi col}) only when $y$ is a small quantity. When $y$ is large, one should instead match onto EFT operators that produce the expanded delta function in Eq. (\ref{eq: expanded delta}). We illustrate this point further with a power counting discussion in Appendix \ref{app: powercounting}, pointing out that it is necessary to count $y$ as small in order to achieve consistent power counting before and after the loop integration in Fig. \ref{fig: Delta loops}. 

The desired operators for the large $y$ regime can be produced by taking the $yP^+ \gg m_\pi$ limit in the definition of the splitting function, expanding away the small momentum in the operator. Such a procedure would give,
\begin{equation}
    f_{\pi p}(y) = \int \frac{db^-}{2\pi} e^{-iy P^+b^-} \bra{p} {\cal O}_\pi(0,0) \ket{p},
\label{eq: expanded fpi}
\end{equation}
where ${\cal O}_\pi(0,0)$ is given by Eq. (\ref{eq: Opi}) with $b$ set to 0. We note this procedure is similar to how heavy quark PDFs and fragmentation functions are studied in heavy quark effective theory, where small momenta are expanded away in powers of $k/m_Q$ \cite{Jaffe:1993ie, Fickinger:2016rfd, vonKuk:2024uxe, Dai:2023rvd}.

To match the different operators onto distinct $y$ regimes, we need to separate large and small $y$ regions in integral of Eq. (\ref{eq: col conv}). As a simple prescription we separate regions with some value, $y_{max}$,
\begin{equation}
    f_{qp}(x)= \sum_H \bigg[\int_x^{y_{max}} + \int_{y_{max}}^1 \bigg]\frac{d y}{y} q^{v}_{H} \bigg(\frac{x}{y}\bigg) f_{Hp} (y).
\end{equation}
For small $y$, $y \le y_{max}$, we match onto the original HDFs in Eq. (\ref{eq: fpi col}). However, for $y > y_{max}$ we instead need to match onto the expanded operators in Eq. (\ref{eq: expanded fpi}). In this regime, however, $f_{\pi p} = 0$, so for the $\bar{d}-\bar{u}$ asymmetry , we are left with the modified expression,
\begin{equation}
    \bar{d}_p(x) - \bar{u}_p(x) = \int_x^{y_{max}} \frac{d y}{y} \bigg[ \bar{d}^v_{\pi^+}\bigg(\frac{x}{y}\bigg) - \bar{u}^v_{\pi^+}\bigg(\frac{x}{y}\bigg)\bigg] f_{\pi p} (y).
\end{equation}
We emphasize that the $y_{max}$ prescription is not the same as a cutoff on the pion's momentum in a loop integral. We have already used dimensional regularization to evaluate the loop integrals, integrating over all values of the pion's momentum, $k$. Instead, $y_{max}$ is a parameter separating the momentum regions where different EFT operators are necessary. While it turns out that $f_{\pi p}(y) = 0$ for large $y$, it could have been that $f_{\pi p}$ was non-vanishing for $y > y_{max}$, in which case we would have had additional contributions from the integral between $y_{max} < y < 1$. 

The importance of separating the momentum regions with the $y_{max}$ prescription is illustrated in the second panel of Fig. \ref{fig: col. d-u}, where we now find excellent agreement with the data. In practice, the exact value of $y_{max}$ is somewhat unclear. As a conservative estimate, we vary $y_{max}$ between $0.25 \sim 2m_\pi/M$ and $ 0.4 \sim 3m_\pi/M$ which yields the relatively large uncertainty on our result. In practice, a smaller range of $y_{max}$ could be determined by comparison with data. 

In Fig. \ref{fig: col. d-u}, the importance of the $\Delta$ resonance is also demonstrated. A simple calculation including only the first diagram of Fig. \ref{fig: Delta loops} is given by the dotted blue line and we see that the calculation agrees with data for $x > 0.15$, however it diverges much more rapidly at small $x$. It is only after including the $\Delta$ states, which have a negative overall contribution because of the $\pi^-\Delta^{++}$, that we are able to properly describe the small $x$ data. 

%{\color{red} Discuss that you need to powercount $y$ as $m_\pi/M$ properly to reproduce the naive power counting}

We now move on to a phenomenological study of TMDs in $\chi$PT using the framework introduced in Ref. \cite{Copeland:2024wwm}. In this paper, it was proposed that the most general matching one can write down is given by 
\begin{equation}
\begin{aligned}
    &f_{q/p}(x, \qv_T) = \sum_H \int d^2 \pv_T d^2 \kv_T \bigg[\int_x^{y_{max}} + \int_{y_{max}}^1\bigg]\frac{d y}{y}\\
    &\times  q^{v}_{H} \bigg(\frac{x}{y}, \pv_T\bigg)  f_{Hp} (x, \kv_T)\delta^{(2)}(\pv_T + \kv_T - \qv_T).
\end{aligned}
\label{eq: gen TMD matching}
\end{equation}
where now the TMD PDFs are matched onto the low-energy TMD HDFs. Notice, for the same reasons as above, we have separated the regions of integration over $y$. The TMD HDFs are defined similar to their collinear counterparts in Eq. (\ref{eq: fpi col}), but now the operators have intrinsic transverse momentum, $\kv_T$. For example, for $y \le y_{max}$, the TMD HDF for the pion in the proton is given by 
\begin{equation}
\begin{aligned}
    f_{\pi p}(y, \kv_T) & = \int \frac{db^-}{2\pi} \frac{d^2\bv_T}{(2\pi)^2}e^{-iy P^+b^-} e^{-i\kv_T \cdot \bv_T} \\
    &\times\bra{p} {\cal O}_\pi(b,0) \ket{p}
\end{aligned}
\label{eq: fpi TMD}
\end{equation}
where ${\cal O}_\pi(b,0)$ is defined in Eq. (\ref{eq: Opi}), except the fields now have transverse separation, in addition to separation along the lightcone, $b = (b^-, 0, \bv_T)$. For $y > y_{max}$ we instead match the TMD PDF onto,
\begin{equation}
\begin{aligned}
    f_{\pi p}(y, \pv_T) & = \int \frac{d^2\bv_T}{(2\pi)^2} e^{-i\kv_T \cdot \bv_T} \bra{p} {\cal O}_\pi(\bv_T,0) \ket{p} \delta(yP^+)
\end{aligned}
\label{eq: fpi TMD large y}
\end{equation}
which again gives zero since it is applied in a regime where $y > y_{max} >0 $.

In analogy with the collinear formalism, the high-energy coefficients in the matching are identified to be the valence TMD PDFs in the intermediate hadron, $q_H^v(\alpha, \pv_T) = q^{(0)}_{H}(\alpha, \pv_T) - \bar{q}^{(0)}_{H}(\alpha, \pv_T)$, where
\begin{equation}
\begin{aligned}
    &q^{(0)}_{H}(\alpha, \pv_T) =\int \frac{db^-}{2\pi} \frac{d^2\bv_T}{(2\pi)^2}e^{-i\alpha P^+b^-} e^{-i\pv_T \cdot \bv_T} \\
    &\times\bra{H} \overline{\psi_i}(b^-, b_T) W_\sqsubset(b, 0) \slashed{n} \psi_i(0) \ket{H}\big|_{m_\pi, p_\pi = 0}.
\end{aligned}
\end{equation}
Notice that, like its collinear counterpart $q^{(0)}_{H}(y)$, this distribution is defined in the limit that the pion mass and small momenta $p_\pi \sim m_\pi$ are set to zero. This is a statement that $q^{(0)}_{H}(y, \pv_T)$ only contains high energy physics at the scale $\Lambda_\chi \gg m_\pi$.

At face value, Eq. (\ref{eq: gen TMD matching}) doesn't appear to be very predictive because the TMD PDFs in the intermediate hadrons, $q^{(0)}_{H}(y, \pv_T),$ are not well-constrained quantities. However, because of the high-energy nature of these objects we note the transverse momentum carried by these objects should be $\pv_T \sim {\cal O}(\Lambda_\chi$). Therefore, we can expand the $q^{(0)}_{H}(y, \pv_T)$ in Eq. (\ref{eq: gen TMD matching}) in powers of $\Lambda_{QCD}/\Lambda_\chi$ to produce a new expression,
\begin{equation}
\begin{aligned}
    &f_{q/p}(x, \qv_T;\mu) = \sum_H \int_x^{y_{max}} \frac{dy}{y}\\
    &\times q^v_H\bigg(\frac{x}{y};\frac{\mu}{\Lambda_\chi}\bigg)f_{Hp}\bigg(y,\qv_T;\frac{m_\pi}{\Lambda_\chi}\bigg) + {\cal O}\bigg(\frac{\Lambda_{QCD}}{\Lambda_\chi}\bigg).
\end{aligned}
\label{eq: LE TMD matching}
\end{equation}
In this expression, the TMD HDFs are convolved with collinear PDFs in the intermediate hadron, not TMD PDFs. Eq. (\ref{eq: LE TMD matching}) argues that, for total transverse momentum $\qv_T \sim m_\pi \ll \Lambda_\chi$, the TMD HDF carries all of the transverse momentum of the TMD PDF because we can treat $q^{(0)}_{H}(y, \pv_T)$ as a semi-perturbative quantity. Eq. (\ref{eq: LE TMD matching}) is the result we would have produced from the start if we assumed from that the high-energy coefficients couldn't describe small transverse momenta scales \cite{Copeland:2024wwm}, which is reassuring.

Here, we have essentially performed an operator product expansion (OPE) on  $q^{(0)}_{H}(x, \pv_T)$ in order to match it on to the PDF. In Eq. (\ref{eq: LE TMD matching}) we have written only the leading order term, where the matching coefficient is a delta function, however there are also perturbatively calculable corrections to this expression. In general, we can write
        \begin{equation}
        \begin{aligned}
        &f_{q/p}(x, \qv_T) = \sum_{H}  \sum_{\rho = u, d, g..}\int_x^{y_{max}} \frac{d y}{y}\int^{1}_\alpha \frac{d\sigma}{\sigma} \int d^2 \pv_T d^2 \kv_T\\
        &\times C_{q\rho}\bigg(\frac{\alpha}{\sigma}, \pv_T\bigg) \rho_{H}^{{v}}(\sigma) f_{Hp} (y, \kv_T)\delta^{(2)}(\pv_T + \kv_T - \qv_T).
        \label{eq: ME TMD matching}        
        \end{aligned}
        \end{equation}
where $\alpha = x/y$. Here, the fixed order matching coefficients, $ C_{q\rho}$, are the usual coefficients found when matching a TMD PDF onto collinear PDFs using perturbative QCD in the large $\pv_T$ limit \cite{Boussarie:2023izj}. Their dependence on the scale $\mu$ and the rapidity scale $\zeta$ is implicit in Eq. (\ref{eq: ME TMD matching}).

In the opposing limit, when $\qv_T \sim \Lambda_\chi \gg m_\pi$, one can expand {\it both} the TMD PDFs in the intermediate hadron, $q_H^v$ and the TMD HDFs in powers of $\Lambda_{QCD}/\qv_T$, which takes the transverse separation $\bv_T \to 0$ in position space. It is straightforward to show that this produces the collinear convolution formula given by Eq. (\ref{eq: col conv}) at leading order. From here, we can again calculate the perturbative coefficients for matching $q_H^v(\alpha, \pv_T)$ onto $q_H^v(\alpha)$. The result is given by, 
       \begin{equation}
       \begin{aligned}
       &f^a_{q/p}(x, \qv_T) = \sum_{H}  \sum_{\rho = u, d, g,..}  
       \int_x^{y_{max}} \frac{d y}{y}\int^{1}_\alpha \\
       &\times\bigg[C_{q\rho}\bigg(\frac{\alpha}{\sigma}, \qv_T\bigg) \rho_{H}^{{v}}(\sigma) f^a_{Hp} (y)+{\cal O}\bigg(\frac{\Lambda_{QCD}^2}{\qv_T^2}\bigg)\bigg].
      \label{eq: HE TMD matching}        
      \end{aligned}
      \end{equation}
where, again, the coefficients can be found Ref. \cite{Boussarie:2023izj}. In this expression, all of the transverse momentum dependence comes from the pQCD matching coefficients.
%\begin{figure*}[htp]
\begin{figure}[htp]
    \centering
    \includegraphics[width=\linewidth]{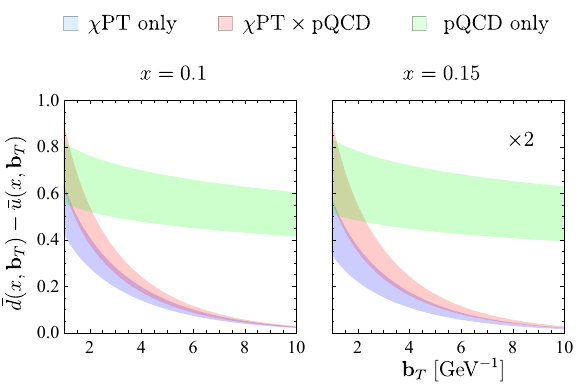}
    \vspace{-.35cm}
    \caption{Different matching prescriptions proposed for the $\bar{d}(x,\bv_T) -\bar{u}(x,\bv_T)$ asymmetry in the proton. The curves represent Eq. (\ref{eq: LE TMD matching}) (blue), Eq. (\ref{eq: ME TMD matching}) (red), and Eq. (\ref{eq: HE TMD matching}) (green). The uncertainty on the bands are from varying $0.3 \le y_{max} \le 0.35$. The pion PDFs and matching coefficients are evaluated at $\mu^2 = \zeta^2 = 54$ GeV$^2$, where $\zeta$ is the rapidity scale.}
    \label{fig: OPE matchings}
    \vspace{-.25cm}
\end{figure}

A major point of this paper is to compare and contrast each matching prescription through an exploratory study of the $\bar{d}-\bar{u}$ asymmetry in the proton's TMD PDFs. For numerical simplicity, we Fourier transform Eqs. (\ref{eq: gen TMD matching}) and (\ref{eq: LE TMD matching}) $-$ (\ref{eq: HE TMD matching}) into $\bv_T$ space. In position space, the convolutions between the valence TMD PDFs and the TMD HDFs turn into products. For example, Eq. (\ref{eq: gen TMD matching}) becomes,
\begin{equation}
    \tilde{f}_{i/p}(x, \bv_T) = \sum_H \int_x^{y_{max}} \frac{d y}{y} \tilde{q}^{v}_{H} \bigg(\frac{x}{y}, \bv_T\bigg)  \tilde{f}_{Hp} (x, \bv_T)
\
\end{equation}
and similar expressions follow for Eqs. (\ref{eq: LE TMD matching})$-$(\ref{eq: HE TMD matching}). 

As in the collinear case, for the TMD $\bar{d}-\bar{u}$ asymmetry we only need the contribution from the $\tilde{f}_{\pi p}(y, \bv_T)$ TMD HDF in the proton, defined in Eq. (\ref{eq: fpi TMD}). Likewise, we only need the $\bar{d}-\bar{u}$ valence TMD PDFs in the intermediate pions as input. Now, as already mentioned, the pion TMD PDFs are not well constrained making a basic application of a convolution like Eq. (\ref{eq: gen TMD matching}) impossible without introducing some sort of model for the pion TMDs. In Eqs. (\ref{eq: LE TMD matching})$-$(\ref{eq: HE TMD matching}), however, the pion TMDs are replaced with the collinear pion PDFs. Therefore, we are able to implement these matching prescriptions without introducing any further model dependence. 

In Fig. \ref{fig: OPE matchings} we plot Eqs. (\ref{eq: LE TMD matching}), (\ref{eq: ME TMD matching}), and (\ref{eq: HE TMD matching}), labeled ``$\chi$PT only", ``$\chi$PT$\times$pQCD", and ``pQCD only", respectively. Again, we use the $y_{max}$ prescription but now we vary $y_{max}$ between $0.3$ and $ 0.35$ because this range of $y_{max}$ describes the collinear data in Fig. \ref{fig: col. d-u} fairly well. With this small range of $y_{max}$, and by neglecting other sources of error, we have almost certainly underestimated the uncertainty on our calculations, however the qualitative features of Fig. \ref{fig: OPE matchings} are not affected by our uncertainties (or lack thereof). We show our expressions for $x = 0.1$ and $x = 0.15$, since the asymmetry is most relevant at low $x$ and the data between $0.1 \le x \le 0.15$ in Fig. \ref{fig: col. d-u} is well described by the chosen $y_{max}$ range. We also plot for $\bv_T > 2$ GeV$^{-1}$, where the long-distance effects of the effective theory are most valid. We include contributions from the $p\to \pi^+n$ and $p\to \pi \Delta$ splittings. Analytic results are given in Appendix \ref{app: HDF results}. One main takeaway from Fig. \ref{fig: OPE matchings} is that including the transverse momentum dependence from $\chi$PT (as done in the blue and red curves) introduces a substantial suppression as $\bv_T$ gets large. On the other hand, the green ``pQCD only" result, which only has transverse momentum dependence from the pQCD matching coefficients, is never really suppressed. Analytically, this occurs because we find the chiral TMD HDFs scale like $\sim K_0(\bv_T m_\pi)$ as $\bv_T$ gets large, where $K_0$ is a modified Bessel function, giving exponential suppression to our answers. Many phenomenological studies of TMDs introduce ad hoc models or parametrizations (justified as additional nonperturbative effects) in order to induce such an exponential suppression \cite{Boussarie:2023izj,Signori:2013mda, Anselmino:2013lza,Scimemi:2019cmh,Bertone:2019nxa, Bacchetta:2022awv, Moos:2023yfa, Bacchetta:2024qre}, so it is interesting that the $\chi$PT calculation generates this qualitative feature naturally.

\begin{figure}[htp]
    \centering
    \includegraphics[width=\linewidth]{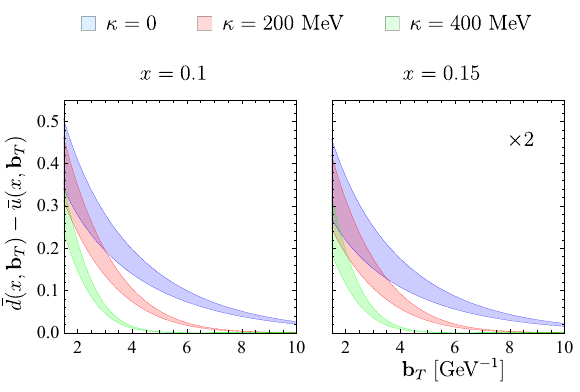}
    \vspace{-.25cm}
    \caption{The $\bar{d}(x, \bv_T)-\bar{u}(x, \bv_T)$ with a Gaussian model for the pion TMD PDFs, given by Eq. (\ref{eq: gaussian model}). The blue curve approximates the pion TMD PDF to match onto the collinear PDF at leading order by setting $\kappa$=0. The red and green curves model the pion TMD PDF with $\kappa = 200$ MeV and $\kappa = 400$ MeV, respectively. The uncertainty comes from varying $0.3 \le y_{max} \le 0.35$. The pion PDFs are evaluated at $\mu^2 = 54$ GeV$^2$.}
    \label{fig: gaussian model} 
\end{figure}

The validity of equations Eq. (\ref{eq: LE TMD matching}) and (\ref{eq: ME TMD matching}) relies heavily on the assumption that no additional nonperturbative behavior lies in the valence TMD PDFs of the intermediate hadron. It is possible however that there is additional physics at the scale $\Lambda_{QCD}$ that $\chi$PT cannot describe, say for example, confining effects for the quark in the intermediate hadron. These effects may be encoded in $\tilde{q}_H^v(x, \bv_T)$ and show up through some additional $\bv_T$ dependence in the distribution because the expansion in $\Lambda_{QCD}/\Lambda_\chi$ in Eq. (\ref{eq: LE TMD matching}) is not a good one. As a simple parametrization, we describe this possible additional $\bv_T$ dependence with,
\begin{equation}
    \tilde{q}_H^v(x, \bv_T) = q^{v}_{H}(x) e^{-\kappa^2 \bv_T^2}.
\label{eq: gaussian model}
\end{equation}
Such a model is also used in studies of TMD quantities in other EFTs, such as NRQCD and HQET \cite{vonKuk:2023jfd,vonKuk:2024uxe,Copeland:2023qed,Copeland:2023wbu}. Here, we vary the parameter $\kappa$ from $\kappa = 200$ MeV $\sim \Lambda_{QCD}$ to $\kappa = 400$ MeV. We only compare this model against the matching prescription proposed in Eq. (\ref{eq: LE TMD matching}) (which can be thought of as setting $\kappa$ = 0) because we are contrasting different long distance effects.

In Fig. (\ref{fig: gaussian model}) we show our results and find a dramatic difference between the curves in the large $\bv_T$ regime. Of course, this could have been expected analytically because adding in a Gaussian with Eq. (\ref{eq: gaussian model}) introduces additional suppression as $\bv_T \to 1/\Lambda_{QCD} \sim 5$ GeV$^{[-1]}$. Qualitatively, however, such a result is interesting because it shows an avenue to discern the importance of additional nonperturbative effects in the valence TMD PDFs in the pion. If the chiral formalism should be sensitive to the $\bv_T$ dependence from the pion TMD PDFs then this will be visible in a comparison against data. Such a scenario would demonstrate that only the general matching prescription in Eq. (\ref{eq: gen TMD matching}) is valid and further work is necessary to disentangle the short and long distance $\bv_T$ behavior of the TMD PDF. On the other hand, it is possible that a measurement of the TMD $\bar{d}-\bar{u}$ asymmetry does not fall off as rapidly and the long-distance $\bv_T$ data is well described by Eq. ($\ref{eq: LE TMD matching}$), where all of the large $\bv_T$ dependence is encoded by chiral TMD HDF. This would imply our formalism for matching TMD PDFs onto $\chi$PT provides a predictive framework to study long-distance physics in TMDs. Such a result would immediately motivate phenomenological studies for the other TMD PDFs in the proton using $\chi$PT.

In this work we have initiated the first phenomenological study of unpolarized TMDs in $\chi$PT. This analysis motivates the experimental measurement of the $\bar{d}-\bar{u}$ asymmetry in the proton TMD PDFs as a key probe into different short and long-distance effects in TMDs. In particular, such a measurement will highlight the possible importance of confinement effects in the TMD PDF and will greatly illuminate our understanding of various nonperturbative mechanisms in the proton. Moving forward, it will be necessary to extend the application of $\chi$PT to other TMD observables, such as other sea quark TMD PDFs in proton and even the valence quark TMD PDFs as well. For example, it should be straightforward to extend SU(3) $\chi$PT calculations of a $s-\bar{s}$ asymmetry \cite{Wang:2016ndh, Wang:2016eoq, Salamu:2019dok} to the TMD PDFs as well. Additionally, while a $\chi$PT investigation of the individual $u$ or $d$ TMDs in the proton will be more complicated, such an analysis should, in principle, be feasible. These studies should be conducted in future work. 

\acknowledgements The authors thank Berndt Mueller, Wally Melnitchouk, Chueng Ji,  Iain Stewart, Duff Neill, Chris Lee, Ivan Vitev and Emanuele Mereghetti for insightful conversations. M.C. thanks Los Alamos National Laboratories for their hospitality while some of this work was completed. M.~C. and T.~M. are supported by the U.S. Department of Energy, Office of Science, Office of Nuclear Physics under grant contract Numbers  DE-FG02-05ER41367. T.~M. is also supported by  the Topical Collaboration in Nuclear Theory on Heavy-Flavor Theory (HEFTY) for QCD Matter under award no.~DE-SC0023547. M.C. is supported by the National Science Foundation Graduate Research Fellowship under Grant No.~DGE 2139754.

\bibliography{main}

\newpage
%%%%%%%%%%%%%%%%%%%%%%%%%
\clearpage
\pagebreak
\pagebreak
\onecolumngrid
\appendix

% ............................................................................
\section{Hadronic distribution function results}
\label{app: HDF results}

In this section we list the results for the $\chi$PT calculations of the collinear and TMD HDFs. Additionally, we provide a useful integral used for the Fourier transform TMD result to $\bv_T$ space. 
%
%%%%%%%%%%%%%%%%%%%%%%%%%%%%
\subsection{Collinear HDFs}
%%%%%%%%%%%%%%%%%%%%%%%%%%%%
In this section, we list the results for the collinear $f_{\pi p}(y)$, which receives contributions from the diagrams in Fig. \ref{fig: Delta loops}. We use the notation $f_{\phi B p}$ to denote the contributions from different intermediate states, where $\phi$ indicates the meson and $B$ the intermediate baryon. At leading order in SU(2) $\chi$PT we have \cite{Salamu:2014pka},
\begin{equation}
    f_{\pi p}(y) = f_{\pi^+n p}(y)-f_{\pi^-\Delta^{++} p}(y)+f_{\pi^+\Delta^0 p}(y) 
\end{equation}
which correspond to the diagrams in Fig. \ref{fig: Delta loops}. The $\pi^-\Delta^{++}$ contribution is negative because this HDF probes the $\pi^-$ valence distribution instead of the $\pi^+$. The unintegrated results for these contributions have already been presented in Ref. \cite{Salamu:2014pka}, we merely present the answers after integrating over $\kv_T$ using dimensional regularization and the MS scheme.

We find the leading contribution to $f_{\pi^+np}(y)$ is given by,
\begin{equation}
    f_{\pi^+n p}(y) = \frac{-2 g_A^2 M^2}{(4\pi f_\pi)^2} y \bigg[\frac{m_\pi^2(1-y)}{m_\pi^2(1-y)+ M^2 y^2} + \log\bigg(\frac{m_\pi^2(1-y)+M^2y^2}{\mu_\chi^2}\bigg)\bigg].
\label{eq: npi col}
\end{equation}
For the numerical calculations in Fig. \ref{fig: col. d-u} we use $g_A = 1.267$ and $f_\pi = 0.093$ GeV. We pick the natural choice for the scale, $\mu_\chi = \Lambda_\chi = 4\pi f_\pi \sim 1.2$ GeV.

Technically, this diagram also receives a contribution from the endpoint, $y\to 0$ \cite{Burkardt:2012hk, Salamu:2014pka}, however this contribution doesn't appear in our numerical comparisons against the E866 and E906 data, which are at nonzero values of the momentum fraction, $x$. Therefore we do not show this contribution. These endpoint terms are relevant, however, when comparing against the integrated asymmetry, $\int_0^1dx (\bar{d}(x) - \bar{u}(x))$.

Using the leading order $\pi N \Delta$ interaction Lagrangian \cite{Scherer:2012xha}, we can compute the contributions from the $\Delta$ resonance as well. We find these contributions are given by $f_{\pi^-\Delta^{++} p}(y) = 3f_{\pi^+\Delta^0 p}(y)$ where
\begin{equation}
\begin{aligned}
    f_{\pi^+\Delta^0 p}(y) = & \frac{g_{N\Delta}^2 (\overline{M}^2-m_\pi^2)}{18f_\pi^2(4\pi M_\Delta)^2} y \bigg[\frac{(\overline{M}^2-m_\pi^2)(1-y)(\delta^2-m_\pi^2)}{m_\pi^2(1-y) + y(M_\Delta^2 -M^2(1-y))}\\
    &- (4M M_\Delta + 3(\delta^2 -m_\pi^2))\log\bigg(\frac{m_\pi^2(1-y) + y(M_\Delta^2 -M^2(1-y))}{\mu_\chi^2}\bigg)\bigg]
\end{aligned}
\end{equation}
where $\overline{M} = M_\Delta + M$ and $\delta =  M_\Delta - M$. Here the coupling is given by $g_{N\Delta} = 3\sqrt{2}g_A/5$ and again we have evaluated the integrals using dimensional regularization and the MS scheme. We note $f_{\pi^+\Delta^0 p}(y)$ also receives contributions from the endpoints $y\to 0$ and $y \to 1$, however the $y\to 0$ region doesn't contribute to our analysis for the same reasons as mentioned above and the $y\to 1$ region is omitted because of the $y_{max}$ prescription.

\subsection{TMD HDFs}

Here we list the results for the TMD HDFs. Like the collinear HDFs, the contributions to $f_{\pi p}(y, \kv_T)$ can be decomposed into three pieces.
\begin{equation}
    f_{\pi p}(y, \kv_T) = f_{\pi^+n p}(y, \kv_T)-f_{\pi^-\Delta^{++} p}(y, \kv_T)+f_{\pi^+\Delta^0 p}(y, \kv_T) 
\end{equation}
The dominant contribution comes from $f_{\pi^+n p}(y, \kv_T)$, which was calculated in Ref. \cite{Copeland:2024wwm}. Here we simply list the result,
\begin{equation}
\begin{aligned}
    f_{\pi^+ n p}(y, \qv_T) =& \frac{g_A^2}{8\pi^3f_\pi^2}\frac{M^2 y(\qv_T^2+M^2y^2)}{[\qv_T^2+M^2y^2+m_\pi^2(1-y)]^2}.
\end{aligned}
\label{eq: final fpiN}
\end{equation}
where we use the same values for $g_A$ and $f_\pi$ as above. We have also dropped the contribution from the $y = 0$ endpoint for the same reasons as before. Notice that the result is regulator parameter and scale independent. This is because the integrals for this calculation were completely convergent so no regularization prescription was needed to obtain the result. 

As in the collinear case, we find the contributions from the $\Delta$ resonance to be $f_{\pi^-\Delta^{++} p}(y, \kv_T) = 3f_{\pi^+\Delta^0 p}(y, \kv_T) $ where

\begin{equation}
\begin{aligned}
    f_{\pi^+\Delta^0 p}(y, \kv_T) = & \frac{g_{N\Delta}^2 (\overline{M}^2-m_\pi^2)}{18f_\pi^2(4\pi M_\Delta)^2\pi} y  \bigg[\frac{(\overline{M}^2-m_\pi^2)(1-y)(\delta^2-m_\pi^2)}{(\kv_T^2+m_\pi^2(1-y)+M_\Delta^2y - M^2(1-y)y)^2}\\
    &+ \frac{ 4M M_\Delta + 3(\delta^2 -m_\pi^2)}{\kv_T^2+m_\pi^2(1-y)+M_\Delta^2y - M^2(1-y)y}\bigg]
\end{aligned}
\end{equation}
where, as above, we have dropped the endpoint contributions at $y =0$ and $y=1$.

We can Fourier transform our results to find the answers in $\bv_T$ space $\tilde{f}_{\pi p}(y, \bv_T) = \tilde{f}_{\pi^+n p}(y, \bv_T)-\tilde{f}_{\pi^-\Delta^{++} p}(y, \bv_T)+\tilde{f}_{\pi^+\Delta^0 p}(y, \bv_T) $. Using the integral \cite{Echevarria:2020qjk},
\begin{equation}
    F_n(\bv_T, \Delta_B) = \int d^{2-2\epsilon}\kv_T \frac{e^{i\kv_T \bv_T}}{(\kv_T^2+\Delta_B)^{n+1}} = \frac{2\pi^{1-\epsilon}}{\Gamma[n+1]}\bigg(\frac{\bv_T^2}{4\Delta_B}\bigg)^{(n+\epsilon)/2}K_{-n-\epsilon}\big(\sqrt{\bv_T^2 \Delta_B}\big)
\label{eq: Fn int}
\end{equation}
where $K_{-n-\epsilon}$ is a modified Bessel function, we find the $\pi^+n$ contribution to be
\begin{equation}
    \tilde{f}_{\pi^+np}(y,\bv_T) = \frac{g_A^2M^2 y}{8\pi^3f_\pi^2}\bigg[F_0(\bv_T, \Delta_N) - m_\pi^2(1-y) F_1(\bv_T, \Delta_N)\bigg]
\end{equation}
where $\Delta_N = M^2y^2+m_\pi^2(1-y)$ and the $\Delta$ contributions to be 
\begin{equation}
    \tilde{f}_{\pi^+\Delta^0 p}(y, \bv_T) = \frac{g_{N\Delta}^2 (\overline{M}^2-m_\pi^2)}{18f_\pi^2(4\pi M_\Delta)^2\pi} y  \bigg[(\overline{M}^2-m_\pi^2)(1-y)(\delta^2-m_\pi^2)F_1(\bv_T, \Delta_\Delta)+ (4M M_\Delta + 3(\delta^2 -m_\pi^2))F_0(\bv_T, \Delta_\Delta)\bigg]
\end{equation}
where $\Delta_\Delta = m_\pi^2(1-y)+M_\Delta^2y - M^2(1-y)y$. Since, numerically, we plot $\tilde{f}_{\pi p}(y, \bv_T)$ away from $\bv_T = 0$,
$F_n(\bv_T, \Delta_M)$ is convergent so in practice we use $d= 2$ dimensions and set $\epsilon=0$ in Eq. (\ref{eq: Fn int}).

%%%%%%%%%%%%%%%%%%%%%%%%%%%%%%%%%%%%%%%%%%%%%%%%%%%%%%%%%%%%%%%%%%%
\section{Power-counting the pion's momentum fraction}
\label{app: powercounting}
%%%%%%%%%%%%%%%%%%%%%%%%%%%%%%%%%%%%%%%%%%%%%%%%%%%%%%%%%%%%%%%%%%%

As explained in the text above, our results crucially depend on restricting the convolution formalism to regions where the pion's momentum fraction is counted as $ y \sim Q/M$. In this section, we provide another argument for why the pion's momentum fraction, $y$, should be power-counted as $Q/M$ when using the definition in Eq. (\ref{eq: fpi col}). The argument can be summarized as follows: $y$ must be counted as small to produce consistent power counting before and after the pion loop integration.

Take for example, the contributions from the $p \to \pi^+n$ intermediate states. The Feynman diagram produces the result \cite{Burkardt:2012hk, Copeland:2024wwm}
\begin{equation}
\begin{aligned}
    f_{\pi^+np}(y) =& \frac{ g_A^2}{2f_\pi^2}\int \frac{d^4k}{(2\pi)^4}\overline{u}(P)\slashed{k}\gamma_5 \frac{i(\slashed{P}-\slashed{k}+M)}{(P-k)^2-M^2+i\epsilon} k^+ \bigg(\frac{i}{k^2-m_\pi^2+i\epsilon}\bigg)^2\gamma_5 \slashed{k} u(P)\delta(k^+ -yP^+).
\end{aligned}
\end{equation}
If we take the pion's momentum $k$ to be soft and scale like some small scale, $Q \sim m_\pi$, then each pion propagator scales like $Q^{-2}$, each nucleon propagator goes like $Q^{-1}$, each vertex gives a power of $Q$, and the integration measure scales like $Q^4$. Additionally, the operator definition in Eq. (\ref{eq: fpi col}) produces an extra factor of $k^+ \delta(k^+-yP^+) \sim 1$ since $k^+ \sim Q$ and $\delta(k^+-yP^+) \sim Q^{-1}$. We also point out that each relativistic spinor  is normalized so that $u(P) \sim \sqrt{M}$. Therefore, power counting arguments suggest that the leading contribution to the $p \to \pi^+n$ splitting should scale like $M\times Q$.

Now, looking at the answer after integration, given by Eq. (\ref{eq: npi col}) we can perform a similar analysis. Here, $m_\pi$ is a small quantity that goes like $Q$ and $M$ is large. If we simply ignore $y$ in the power counting then we find Eq. (\ref{eq: npi col}) scales like $M^2$ instead of the expected $M \times Q$, which is clearly incorrect. In order to reproduce the correct scaling, we must count $y \sim Q/M$. It is easy to check that this choice for the $y$ scaling enables Eq. (\ref{eq: npi col}) $\sim M \times Q$. Similar arguments hold for the $\Delta$ contributions and for the scaling of the TMD HDFs. 

Interestingly, we note that while our formalism produces the correct power of $M \times Q$ at leading order, our expressions are not uniform in the power counting and include subleading $M \times Q^n$ terms as well. This is because we use a relativistic $\chi$PT formalism which mixes large (${\cal O}(M)$) and small (${\cal O}(Q)$) momentum scales due to the Feynman rule for the baryon propagator. A homogeneous power counting can be achieved by using heavy baryon chiral perturbation theory  (HB $\chi$PT) instead \cite{Jenkins:1990jv}. 

For example, in HB $\chi$PT, we find that the contribution to $f_{\pi^+np}(y)$ is given by,
\begin{equation}
\begin{aligned}
    f_{\pi^+np}(y) =& \frac{ g_A^2M}{2f_\pi^2}\int \frac{d^4k}{(2\pi)^4}\overline{u}_v(P)S_v \cdot k \frac{i}{2v\cdot k +i\epsilon} k^+ \bigg(\frac{i}{k^2-m_\pi^2+i\epsilon}\bigg)^2 S_v \cdot k  ~u_v(P)\delta(k^+ - yP^+)\, ,
\end{aligned}
\end{equation}
where $v^\mu$ is the proton's four-velocity, $u_v(P)$ is a nonrelativistic spinor, and $S_v^\mu$ is the spin operator. We find,
\begin{equation}
    f^{HB}_{\pi^+n p}(y) = \frac{-2 g_A^2 M^2}{(4\pi f_\pi)^2} y \bigg[\frac{m_\pi^2}{m_\pi^2+ M^2 y^2} + \log\bigg(\frac{m_\pi^2+M^2y^2}{\mu_{\chi}^2}\bigg)\bigg].
\label{eq: HB fpin}
\end{equation}
which is exactly the result in Eq. (\ref{eq: npi col}), except without the factors of $ym_\pi^2$. If you again count $y$ as ${\cal O}(Q/M)$ then this result scales uniformly like $M \times Q$, as desired. Since the heavy baryon formalism  naturally drops the $ym_\pi^2$ factors, which are suppressed by an additional factor of $Q$, this is another indication that $y$ must be treated as $O(Q/M)$ in the power counting. We point out that the numerical differences between the HB$\chi$PT result in Eq. (\ref{eq: HB fpin}) and our relativistic calculations in Eq. (\ref{eq: npi col}) are small and do not change the conclusions of our analysis.

%%%%%%%%%%%%%%%%%%%%%%%%%%%%%%%%%%%%%%%%%%%%%%%%%%%

\end{document}